\documentclass{nature}
\usepackage{amsmath}
\usepackage{amssymb}
\usepackage{amsfonts}
\usepackage{graphicx}

\newcommand{\ket}[1]{\left|#1\right\rangle}

\title{Landau-Zener interferometry with superconducting qubits}

\author{Mika Sillanp\"a\"a$^1$, Teijo Lehtinen$^1$, Antti Paila$^1$, Yuriy
  Makhlin$^{2,1}$, \& Pertti Hakonen$^1$}
\begin{document}
\maketitle
\begin{affiliations}
\item Low Temperature Laboratory, Helsinki University of
Technology, FIN-02015 HUT, Finland \item Landau Institute for
Theoretical Physics, Kosygin st. 2, 119334 Moscow, Russia
\end{affiliations}

\begin{abstract}
Quantum-mechanical systems having two discrete energy levels are
ubiquitous in nature. For crossing energy levels, depending on how
fast they approach each other, there is a possibility of a
transition between them. This phenomenon is known as Landau-Zener
tunneling~\cite{Landau32,Zener32,Stueckelberg32,Majorana32,LandauI}
and it forms the physical basis of the Zener diode, for example.
The traditional treatment of the Landau-Zener tunneling, however,
ignores quantum-mechanical interference. Here we report an
observation of phase-sensitive interference between consecutive
Landau-Zener tunneling attempts in an artificial two-level system
formed by a Cooper-pair-box qubit~\cite{Nakamura99,shytovEPJB03}.
We interpret the experiment in terms of a multi-pass analog to the
well-known optical Mach-Zehnder interferometer. In our case, the
beam splitting occurs by Landau-Zener tunneling at the charge
degeneracy, while the arms of the Mach-Zehnder interferometer in
energy space are represented by the ground and excited state. Our
Landau-Zener interferometer can be used as a high-resolution
detector for phase and charge owing to interferometric
sensitivity-enhancement. The findings also demonstrate new methods
for qubit manipulations.
\end{abstract}

Landau-Zener (LZ) tunneling is a celebrated quantum-mechanical
phenomenon, taking place at the intersection of two energy levels
that repel each other due to a weak interaction~\cite{Wigner}. The
LZ theory, developed in the early 1930's in the context of slow
atomic collisions~\cite{Landau32,Zener32,Stueckelberg32} and spin
dynamics in time-dependent fields~\cite{Majorana32}, demonstrated
that transitions are possible between two approaching levels as a
control parameter is swept across the point of minimum energy
separation. The LZ tunneling is often used as a tool for
determining level separations, for example, in molecular
NMR~\cite{magnetism}. The probability of a Landau-Zener tunneling
transition is given
by~\cite{Landau32,Zener32,Stueckelberg32,Majorana32}:
\begin{equation}\label{LZS}
    P_{\mathrm{LZ}}=\exp\left(-2 \pi \frac{\Delta^{2}}{\hbar v}\right)
\end{equation}
where $v\equiv d(\varepsilon_1 - \varepsilon_0)/dt$ denotes the
variation rate of the energy spacing for noninteracting levels,
and $2 \Delta$ is the minimal energy gap.

Yet for quantum-mechanical systems, more fundamental is the
transition \emph{amplitude}, which allows one to describe
interference. As two colliding atoms approach each other, their
electronic levels may cross. The probability amplitudes evolve
along either of the two potential curves and may interfere, when
the levels cross again after the collision. The wave-function
phase accumulated between the incoming and outgoing traversals
varies with the collision energy giving rise to Stueckelberg
oscillations, observed in atomic systems \cite{StuckOsc69}, in the
populations. Typically, however, the phase is large and rapidly
varies with energy, which allows one to average over these fast
oscillations~\cite{LandauI,Stueckelberg32}, neglecting the
interference.

Recently, quantum coherence in mesoscopic Josephson tunnel
junctions has been investigated
extensively~\cite{Nakamura99,MSS,jena}, since they might provide a
realistic platform for quantum-information processing. In these
artificial quantum systems, energy scales can easily be tuned into
a range feasible for study of fundamental phenomena. We used a
system of such mesoscopic junctions to obtain the first evidence
of quantum interference associated to Landau-Zener tunneling in
non-atomic systems.

\begin{figure}
\center
\includegraphics[width=9.0cm]{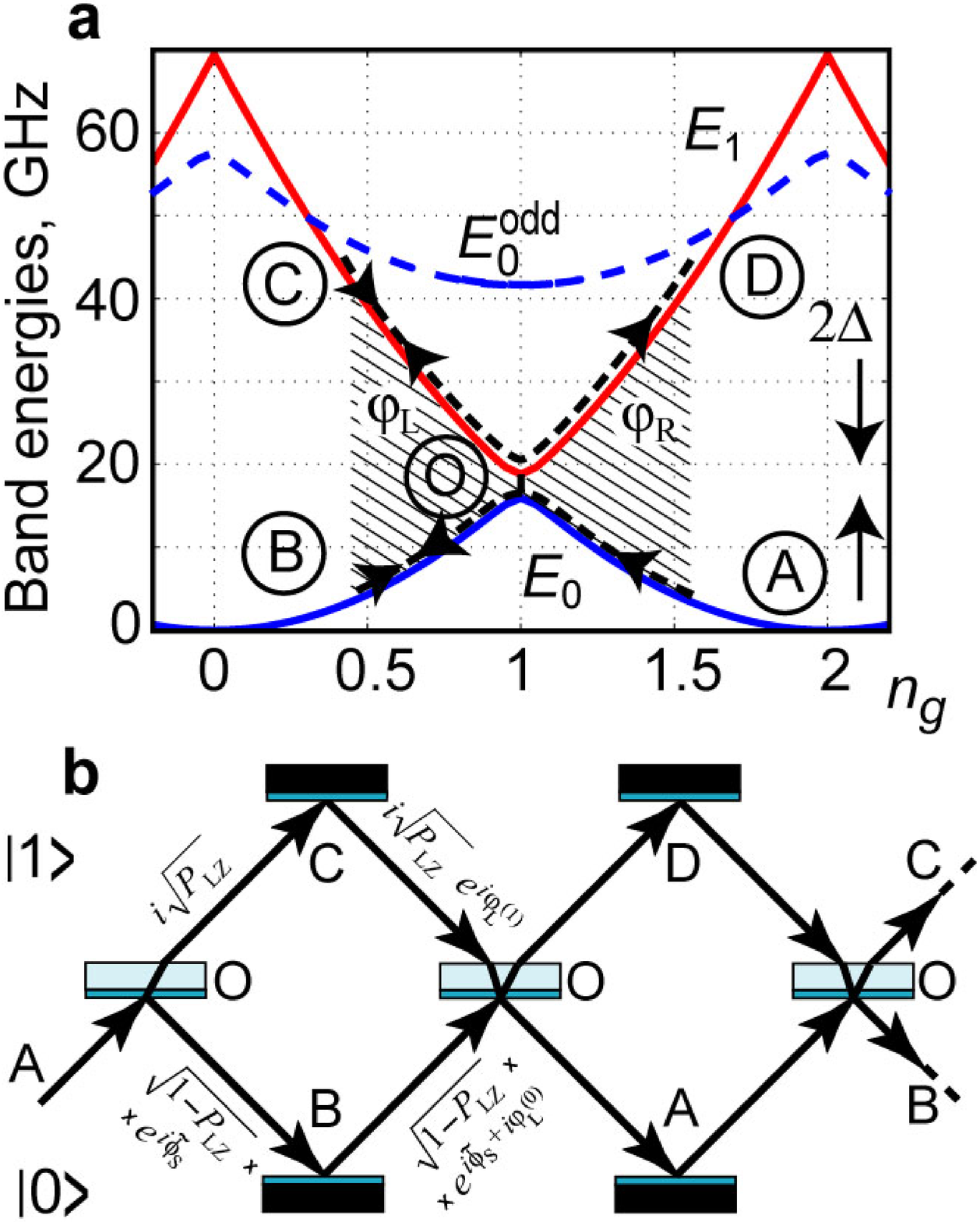}

\caption{{\large\textbf{a}}, schematic view of interference of
successive Landau-Zener (LZ) tunneling attempts in Cooper-pair box
(CPB). From the initial state at A, the state evolves through the
first LZ event at the avoided level crossing at $n_g = 1$ (O)
either towards B (no LZ tunneling) or C (with LZ tunneling). After
the turning points B and C, the final state D is reached either by
a second LZ tunneling or by remaining on the excited band,
respectively. The hatched area determines the dynamical phase
shifts $\varphi_{L,R}$. The level spacing $E_1 - E_0$ varies
roughly as $4E_C(1-n_g)$, \emph{i.e.,} linearly as in the generic
LZ-tunneling problem with linearly crossing energy levels.
$2\Delta_0 \sim 3$ GHz denotes the minimum gap in our experiments.
The dashed line represents the lowest energy of odd parity state
$E_0^{\mathrm{odd}}$. Even states with energy larger than the odd
state value will always try to relax to the odd state.
{\large\textbf{b}}, interpretation of the LZ interference in
\textbf{a} as a multi-pass analog of the optical Mach-Zehnder
interferometer. LZ events correspond to the beam splitters which
divide the wave function into two partial waves, with the
(probability) amplitudes as marked. The mirrors play the role of
the dynamical phases $\varphi^{(0,1)} = \hbar^{-1} \int E_{0,1}
dt$ picked up away from the avoided crossing.} \label{LZbands}
\end{figure}

We used a charge qubit based on a Cooper-pair box (CPB) that we
turned into an analog of the optical Mach-Zehnder interferometer.
In this device, a beam is split into two partial waves, which
interfere after a single passage through the system. In our case,
the LZ tunneling provides the mechanism of the beam splitting and
occurs when the gate charge of the Cooper-pair box is swept across
the degeneracy (Fig.~\ref{LZbands}). The split beams follow the
ground and excited states of the CPB and recombine at the
subsequent degeneracy point. We find a collection of different
types of interference patterns which can be described using the
basic principles of Mach-Zehnder interferometers. Our
interferometric observations are made possible by the non-invasive
character of our dispersive measurement method~\cite{CQED,Mika05}.

\begin{figure}
\center
 \includegraphics[width=12.0cm]{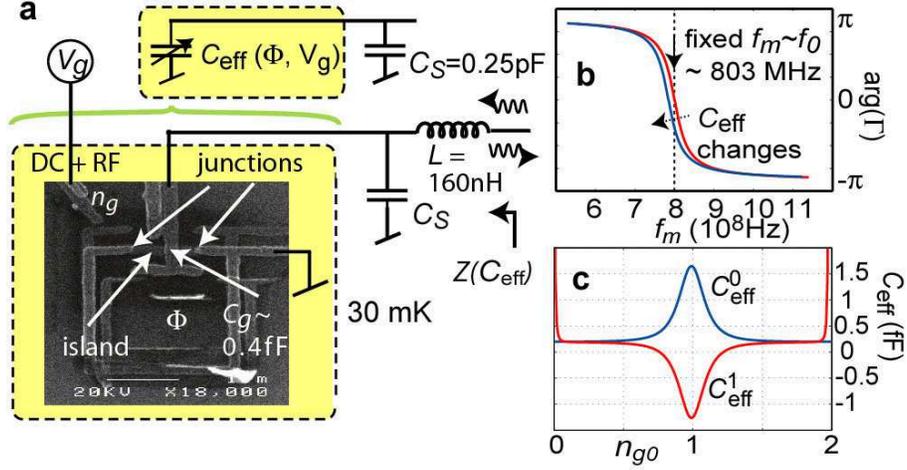}

\caption{{\large\textbf{a}}, schematics of our experiment. The
resonant frequency $f_0$ of the lumped-element $LC$ circuit is
tuned by the Josephson capacitance $C_\mathrm{eff}$ of the CPB
shown in the SEM image. The junctions of the split Cooper-pair box
had an average resistance of $R=23$ k$\Omega$ each, corresponding
a maximum Josephson energy of the box $2E_J=12.5$ GHz, which could
be tuned down to 2.7 GHz by magnetic flux $\Phi$. The capacitance
the junctions amounts to $C_1+C_2\sim$ 0.34 fF, yielding a Coulomb
energy of $e^2/2(C_1+C_2+C_g)=1.1$ K. {\large\textbf{b}},
illustration of the phase shift $\arg (\Gamma)$ of the reflected
microwaves at a fixed measurement frequency $f_m$ while
$C_\mathrm{eff}$ increases. {\large\textbf{c}}, the Josephson
capacitance calculated for the two lowest levels of our CPB with
$E_J/E_C= 0.27$ and asymmetry $d=0.22$ at $\phi =0$. Note that
variations from the "classical" capacitance level, $(1/C_g
+1/(C_1+ C_2))^{-1} \simeq 0.2$ fF, are opposite for the two
levels: ground level capacitance $C_\mathrm{eff}^0 > 0$ while the
excited state has $C_\mathrm{eff}^1 < 0$.} \label{scheme}
\end{figure}

Our superconducting Mach-Zehnder interferometer is made using a
single-Cooper-pair transistor (SCPT) embedded into a small
superconducting loop (Fig.~\ref{scheme}). An SCPT consists of a
mesoscopic island having capacitance $C$, two small Josephson
junctions, and of a nearby gate electrode used to polarize the
island with the (reduced) gate charge $n_g = C_g V_g / e$. The
island has the charging energy $E_C = e^2/(2C) \sim 1$ Kelvin, and
the junctions have the generally unequal Josephson energies $E_J(1
\pm d)$, where $d$ quantifies the asymmetry. With $d=0$, SCPT
Hamiltonian is then $E_C (\hat{\mathrm{n}}-n_g)^2 - 2E_{J} \cos
\left(\phi/2 \right) \cos (\hat{\theta}) - C_g V_g^2/2$. Here, the
number $\hat n$ of extra electron charges on the island is the
quantum conjugate variable to $\hat\theta/2$, where $\hat\theta$
is the superconducting phase on the island. The SCPT is then
equivalent to a Cooper-pair box, but with an effective Josephson
energy of $2 E_J \cos(\phi/2)$ tunable by the superconducting
phase across the two junctions, $\phi = 2 \pi \Phi / \Phi_0$.
Here, $\Phi_0 = h/(2e)$ is the magnetic flux quantum.
Diagonalization of the Hamiltonian leads to energy bands
$E_k(n_g,\phi)$ (supplementary information on bands: see
Ref.~\cite{MIKAthesis}).

When $E_C \gg E_J$, the Hamiltonian is conveniently written in the
eigenbasis $\{| n \rangle \}$ of the charge operator $\hat{n}$,
taking only two charge states into account. Then the Hamiltonian
of the CPB becomes
\begin{equation}\label{Ham}
H=\left(
  \begin{array}{cc}
    \epsilon(n_g) & -\Delta \\
    -\Delta & - \epsilon(n_g) \\
  \end{array} \right)
    \end{equation}
\begin{equation}\label{HamSpin}
= -\frac{1}{2} B_z \sigma_z -\frac{1}{2} B_x \sigma_x
\end{equation}
where $\epsilon =\frac{1}{2} B_z= 2E_C (1-n_g)$ and
$\Delta=\frac{1}{2} B_x=E_J \cos(\phi/2)$.

We display here the Hamiltonian both in the matrix
form~(\ref{Ham}) and in the spin form, Eq.~(\ref{HamSpin}). The
asymmetry $d \neq 0$ in Josephson energies would limit the minimum
value for the off-diagonal coupling $|\Delta|$. The eigenvalues
$E_0(n_g, \phi)$ and $E_1(n_g, \phi)$ are the two lowest bands as
illustrated by Fig.~\ref{LZbands}a. By $\ket{0}$ and $\ket{1}$, we
denote the corresponding wave functions.


We analyze the level crossing process using the energy diagram in
Fig.~\ref{LZbands}a. As $n_g$ is lowered and then increased
(similar to the interatomic distance during a collision), the
system can follow either of two possible paths: AOCOD and AOBOD.
The probability to follow either path is a product of two
independent events: $P_{\mathrm{LZ}}$ for making a transition and
$1-P_{\mathrm{LZ}}$ for staying on the same level, which gives the
total transition probability
\begin{equation}\label{LZStwo}
    P_{\mathrm{AD}}=2 P_{\mathrm{LZ}} (1-P_{\mathrm{LZ}})  \,.
\end{equation}
In this treatment interference between tunnel attempts has been
neglected.

Subsequent LZ tunneling events with time interval $\tau_p$ can
interfere, provided phase coherence is preserved and these events
do not overlap~\cite{GefenRev,kayanuma}, $\tau_z < \tau_p <
\tau_{\mathrm{coh}}$. Here, the time of an LZ-tunneling
event~\cite{Mullen89} is $\tau_z \sim \sqrt{\hbar/v} \cdot
\mathrm{max}(1,\sqrt{\Delta^2/\hbar v})$. In charge qubits, it is
easy to make $\tau_z \ll \tau_{\mathrm{coh}}$ where the coherence
time is $\tau_{\mathrm{coh}}=\min(T_1, T_2)$ with $T_1$ and $T_2$
corresponding to the relaxation and dephasing time, respectively.
For example, by taking $\Delta=2$ GHz and $v=40$ GHz per 1 ns, we
obtain $\tau_z\sim 0.1$ ns, which is well within experimental
reach. The interference of consecutive tunneling attempts can be
viewed as two partial waves, describing the propagation along
either the lowest band or the first excited band. This is similar
to an optical Mach-Zehnder interferometer as illustrated in
Fig.~\ref{LZbands}b.

Away from the crossing region, the eigenstates $\ket{0}$ and
$\ket{1}$ accumulate the dynamical phase
\begin{equation}\label{phase}
    \varphi = \varphi^{(1)} - \varphi^{(0)} = \frac{1}{\hbar} \int \left[ E_1(n_g(t)) -
    E_0(n_g(t)) \right]
    dt\, .
\end{equation}
In addition, each pass of the level crossing results in an LZ
event with probability amplitudes given by
(cf.~Fig.~\ref{LZbands}b)~\cite{GefenRev,kayanuma}:
\begin{equation}\label{amplitudes}
\left(
\begin{array}{c}\ket{0}\\\ket{1}\\\end{array}
\right) \Rightarrow
\left(
\begin{array}{cc}
  \sqrt{1-P_{\mathrm{LZ}}} \exp (i \tilde\phi_S) & i\sqrt{P_{\mathrm{LZ}}} \\
  i\sqrt{P_{\mathrm{LZ}}} & \sqrt{1-P_{\mathrm{LZ}}} \exp (-i \tilde\phi_S ) \\
\end{array}%
\right) \left(
\begin{array}{c}\ket{0}\\\ket{1}\\\end{array}
\right)
\end{equation}
Here, $\tilde\phi_S=\phi_S-\pi/2$, where the Stokes phase $\phi_S$
depends on the adiabaticity parameter $\Delta^2/ \hbar v$
(cf.~Eq.~(\ref{LZS})). In the adiabatic limit, $\phi_S \rightarrow
0$, but in the sudden limit, $\phi_S = \pi/4$.

Let us assume a fast gate charge sweep of the form $n_g(t) =
n_{g0} + \delta n_{\mathrm{rf}} \sin (\omega_{\mathrm{rf}} t)$,
where the constant level $n_{g0}$ means that the sweep is in
general offset from the crossing point. One cycle of continuous
driving in our CPB takes the system twice through the crossing
point and involves two dynamical phase shifts $\varphi_L$ and
$\varphi_R$, on the left and right sides. For a single cycle, we
add the amplitudes along the two branches in Fig.~\ref{LZbands}b,
to find the probability of reaching the point D:
\begin{eqnarray}
    P_\mathrm{AD}=\left| i \sqrt{P_{\mathrm{LZ}}(1-P_{\mathrm{LZ}})}
    \exp{\left[ i(\varphi_\mathrm{L}^{(0)} +
    \tilde\phi_S) \right]}+ i \sqrt{P_{\mathrm{LZ}}(1-P_{\mathrm{LZ}})} \exp{\left[
    i(\varphi_\mathrm{L}^{(1)} - \tilde\phi_S)\right]}\right|^2\nonumber\\
    =2 P_{\mathrm{LZ}} (1-P_{\mathrm{LZ}}) \left[1 + \cos (\varphi_\mathrm{L} - 2
    \tilde\phi_S)\right] \,.\label{LZS2}
\end{eqnarray}
Clearly, the maximum transition probability is reached when the
total phase $\varphi_\mathrm{L} - 2 \tilde\phi_S$ is a multiple of
$2\pi$. Under continuous driving, one obtains a multi-pass
Mach-Zehnder model. In this case it can be shown that the maximum
population of $\ket{1}$ (constructive interference) is reached
when both dynamical phases satisfy the condition mentioned above,
\begin{equation}\label{phasecondition}
    \varphi_\mathrm{L,R} - 2 \tilde\phi_S \: \: \: \mbox{  are multiples of  }  \: \: 2\pi
\end{equation}
For example, in the adiabatic limit, $\varphi_\mathrm{L,R}$ have
to be odd multiples of $\pi$. The resonance conditions in
Eq.~(\ref{phasecondition}) are seen overlayed in
Figs.~\ref{C3d4GHz} and \ref{alpha1}
 (see below) as the black solid and dashed lines.


Our experimental scheme is illustrated in Fig.~\ref{scheme}. The
measurement signal tracks the time average, under a strong LZ
drive, of the Josephson capacitance $C_{\mathrm{eff}} \propto
\frac{\partial^2E(\phi, V_g)}{\partial V_g^2}$, probed at
$f_m=803$ MHz (see Methods). In the first approximation, the
energy $E$ here can be taken as the average energy stored in the
qubit: $E=p_0 E_0 + p_1 E_1$, where the band energies are weighted
by their average populations $p_0$ and $p_1$, respectively. Thus,
we have
\begin{equation}\label{Csimple}
C_{\mathrm{eff}}=p_0 C_{\mathrm{eff}}^0 + p_1 C_{\mathrm{eff}}^1
\, .
\end{equation}
In this way, however, we neglect all the relaxation phenomena that
take place on time scales $1/f_p$: If the relaxation rate
$(T_1)^{-1} \gg f_p$ then changes in $p_0=p_0(V_g)$ have to be
taken into account in the response. As will be seen below this is
in fact crucial in understanding our experimental results.

We have made extensive scans of the reflection coefficient of a
CPB by varying the LZ drive frequency $f_{\mathrm{rf}}=0.1-20$
GHz, and its amplitude $\delta n_{\mathrm{rf}}=0-3$ electrons, as
well as the bias $n_{g0}$ and $\phi$. The Josephson capacitance
deduced from the phase shift of the reflected wave $\arg (\Gamma)$
at $f_{\mathrm{rf}}=4$ GHz when changing $\delta n_{\mathrm{rf}}$
and $n_{g0}$ is illustrated in Fig.~\ref{C3d4GHz}. We observe a
clear interference pattern whose main characteristics agree with
those expected for coherent LZ tunneling: 1) There is an onset of
the interference speckles with a distinct value for $n_{g0}$ where
the rf drive just reaches the avoided crossing, linearly dependent
on the AC drive amplitude; 2) The density of the dots is
proportional to $1/f_{\mathrm{rf}}$ in the direction of $n_{g0} $
as well as $\delta n_{\mathrm{rf}}$, 3) The central part of the
interference patterns displays the curved dot rows, in a similar
fashion as in the overlayed patterns, 4) the pattern loses its
contrast at a certain value towards lowering drive frequency, here
at $f_{\mathrm{rf}} \sim 2$ GHz. Note also that there are
destructive interference dots at high drives, where the qubit
remains basically on the lowest level (cf.~the "coherent
destruction of tunneling"~\cite{destruction}).

The periodicity in $n_{g0}$ is clearly $2e$ at low levels of
rf-excitation. At excitations on the order of $e/2$, there is an
appearance of a shifted, additional pattern, which makes the
signal almost $e$-periodic. The origin of these odd sectors can be
understood by looking at the energy levels displayed in
Fig.~\ref{LZbands}. When the rf-drive brings the system past a
crossing point of $E_1$ and $E_0$, it becomes energetically
favorable to enter an odd particle-number state, resulting in a
shift by $e$ in the interference pattern. The odd states appear to
be rather stable~\cite{aumentado}, and the contrast of the "odd
sectors" is almost as strong (see Fig.~\ref{C3d4GHz}).

According to Eq.~(\ref{phasecondition}), the phase difference,
\begin{equation}\label{diffphase}
    \varphi_- \simeq \varphi_\mathrm{L} -\varphi_\mathrm{R} =
    2\pi\,\frac{4E_C(1-n_{g0})}{\hbar
    \omega_{\mathrm{rf}}},
\end{equation}
is a multiple of $2\pi$ at resonances, implying the location of
the population peaks on the lines of fixed gate bias $n_{g0}$ with
spacings $\Delta n_{g0} =\hbar \omega_{\mathrm{rf}}/(2 E_C)$ (we
approximated the dynamical phase by that for non-interacting
levels). This linear dependence of the spacings on frequency is
illustrated in Fig.~\ref{C3d4GHz}c. The fitted line yields $E_C =
1.1$ K, which is about 25~\% higher than we obtained from the
rf-spectroscopy~\cite{Mika05}.

As usual, interference effects are prone to decoherence and our
interferograms are suitable for studying dephasing and relaxation
in qubits as proposed by Shytov~\emph{et al.}~\cite{shytovEPJB03}.
The suppression of interference in our data at low
$f_{\mathrm{rf}}$ is due to the loss of phase memory over a single
LZ cycle. Indeed, phase fluctuations suppress the contrast of
oscillations in Eq.~(\ref{LZS2}) (see Ref.~\cite{GefenRev} for a
more detailed analysis). This way, we find a quick estimate,
averaged over $n_g$, for the coherence time of our qubit
$\tau_\mathrm{coh}\sim 0.5$ ns.


To account for decoherence in a detailed manner, we solved the
phenomenological Bloch equations~\cite{Bloch46,Makhlin03} which
describe the dynamics of the magnetization
$\mathbf{M}=\langle\mathbf{S}\rangle$ of a pseudospin-$1/2$ (a
two-level system):
\begin{equation}\label{BlochEQ}
\frac{d}{dt}\overrightarrow{M}=-\overrightarrow{B} \times
\overrightarrow{M}
-\frac{1}{T_1}(\overrightarrow{M_{\shortparallel}}-\overrightarrow{M_{\shortparallel}^{eq}})
- \frac{1}{T_2}\overrightarrow{M_{\perp}} \,.
\end{equation}
Here, the pseudomagnetic field is given by Eq.~(\ref{HamSpin}).
The parameters $T_1$ and $T_2$ describe the relaxation of the
$z$-component of magnetization towards equilibrium and the
relaxation of transverse magnetization to zero, respectively.
Assuming that decoherence is dominated by charge noise, we write
\begin{equation}\label{t1}
    \frac{1}{T_1}=\frac{\sin^2 \eta}{2\hbar^2}S_X(\omega= (E_1 - E_0)/\hbar)\,,
\end{equation}
\begin{equation}\label{t2}
    \frac{1}{T_2}=\frac{1}{2}\frac{1}{T_1}+\frac{\cos^2 \eta}{2\hbar^2}S_X(\omega=0)\,.
\end{equation}
Here, the angle $\tan\eta = B_z/B_x$ describes the dependence on
the gate bias, and the voltage fluctuations are involved via
$S_X(\omega)=\left(2e\frac{C_t}{C_J} \right)^2 S_V(\omega)$. For
Ohmic dissipation,
$ 
S_X(\omega)=2 \pi \hbar^2 \alpha \omega \coth \frac{\hbar
\omega}{2k_B T}
$ 
where the coefficient $\alpha$ characterizes both the bath and its
coupling to the qubit: $\alpha=
(1+[C_1+C_2]/C_g)^{-2}\frac{2e^2}{h}R$.
For our sample, $\alpha \sim 10^{-2}$ due to strong coupling to
the environment via the high gate capacitance and parasitic
capacitance in the resonator inductance.

The results of the simulations for a strong driving $B_z= 2E_C
(1-n_{g0} + \delta n_{\mathrm{rf}} \sin(\omega_{\mathrm{rf}} t))$
are illustrated in Fig.~\ref{alpha1}. It displays the population
of the ground state on the $ n_{g0}$ vs. $\delta n_{\mathrm{rf}}$
plane for Ohmic bath with $\alpha =0.04$. The variation of the
level population agrees well to the overlayed interference pattern
obtained using the Mach-Zehnder model, except for the regions near
the edges (the inclined white lines in Fig.~\ref{alpha1}). The
additional structure in this region is a signature of multiphoton
transitions: Under the condition of the $n$-photon resonance
($4E_\mathrm{c}(1-n_{g0}^{(n)})=n\cdot\hbar\omega_\mathrm{rf}$)
the effective coupling in the pseudospin rotating frame
is~\cite{Lopez} $\tilde\Delta=\Delta\cdot J_n(4E_\mathrm{c}\delta
n_\mathrm{rf}/\hbar\omega_\mathrm{rf})$, and from the stationary
solution of the Bloch equations (with gate-independent relaxation
terms) we find the population response~\cite{Abragam} $\propto
T_1T_2\tilde\Delta^2 / (1+T_1T_2\tilde\Delta^2 + ({\textstyle
\frac{1}{\hbar}}E_\mathrm{c} (n_{g0}-n_{g0}^{(n)}) T_2)^2)$. The
Bessel functions describe the onset of the resonance as well as
the interference pattern at higher drives. This expression
indicates that destructive interference occurs at amplitudes
corresponding to zeros of the Bessel functions, and that the
destructive interference spots are sharper, in accordance with the
data.

The range of the capacitance variation on the lower level,
$C_{\mathrm{eff}}^{(0)}\sim 0.2$--1.5~fF, deduced from the data
measured without the LZ drive, are in accordance~\cite{Mika05}
with the ground band curvature, shown in Fig.~\ref{scheme}c. As
the maximum population on the upper level can only reach 50~\%,
the weighted average, Eq.~(\ref{Csimple}), should always be larger
than 0.2 fF (see Fig.~\ref{scheme}). Clearly, there are regions in
Fig.~\ref{C3d4GHz} where the response looks stronger. Hence,
instead of Eq.~(\ref{Csimple}), we have to calculate a time
average $\langle-\frac{\partial^2E}{\partial V_g^2}\rangle$, whose
magnitude can be substantially \emph{increased} by relaxation
phenomena.

Assuming very fast relaxation, \emph{i.e.,} with the populations
tracking the instantaneous AC gate charge $n_g= n_{g0} + \delta
n_{\mathrm{rf}} \sin(\omega_{\mathrm{rf}} t)$, we find a
capacitance of good uniformity, in a similar manner to that of the
measured interference patterns. However, at reasonable values of
parameter $\alpha$, the swing of the capacitance is too large
indicating that the assumption of instantaneous relaxation is too
strong.

With a finite relaxation $T_1$, we have calculated the spin
dynamics having a weak measurement signal of amplitude $\delta
n_{\mathrm{ac}}=C_g\delta V_{\mathrm{ac}}$ on. Then, we can use
the linear-response theory to extract the capacitance. We
calculate the time-dependent expectation value for the effective
charge $\langle Q \rangle (t) =
\mathrm{Tr}(\rho*Q_{\mathrm{eff}})$, where $Q_{\mathrm{eff}} =
-dE/dV_g$, and the density matrix is expressed in the energy
eigenbasis. From $\langle Q \rangle (t)$ we pick up the quadrature
components, $Q_{\omega_{\mathrm{in}}}$ and
$Q_{\omega_{\mathrm{out}}}$, at the measurement frequency. Using
the definition of impedance, we may solve for the capacitance
$C_{\mathrm{eff}}=\frac{Q_{\omega_{\mathrm{in}}}^2+Q_{\omega_{\mathrm{out}}}^2}{Q_{\omega_{\mathrm{in}}}\delta
V_{ac}}$. The resulting capacitance at $f_{\mathrm{rf}}=4$ GHz is
illustrated in Fig.~\ref{trace}. The full swing of the capacitance
over the $ n_{g0}$ vs. $\delta n_{\mathrm{rf}}$ plane is very
sensitive to the parameter $\alpha$. The values $\alpha =0.04$ and
$\delta n_{ac}=0.03$ were taken for the calculation in order to
match the measured pattern. This corresponds at the degeneracy
point to $T_2 \sim 0.5$ ns which is close to the estimates
obtained both from the oscillation contrast, and microwave
spectroscopy. The calculation reproduces the major features of the
measured interferograms: 1) The spacing of the dots, 2) the size
of the capacitance swing, 3) the global uniformity of the pattern,
and 4) minima and maxima on separate arcs (rather than on the same
arc).

Similarly as optical interferometry played a central role in the
development of "photon" physics~\cite{Photon}, solid-state quantum
interferometry~\cite{Sprinzak} may find many applications. We
propose to apply the LZ interferometry for sensitive detection of
phase and charge utilizing mesoscopic Josephson circuits
\cite{lset,leif05,Mika05}, where it brings about a significant
increase in sensitivity. Indeed, the LZ interferometer can be
viewed as integrating phase amplifier for the superconductor phase
$\phi$ across the device. The interferometer transforms tiny
changes of $\phi$ (or magnetic flux $\Phi$) into a huge modulation
of the wave-function phase $\varphi$ by basically integrating the
hatched area in Fig.~\ref{LZbands} \cite{Gorelik}, but at the
expense of reduced measurement strength. Eventually, it amounts at
least to $\sim 2\pi$-fold increase in detector sensitivity, with
significantly reduced disturbance due to the measurement signal.
Another possibility would be to separate the LZ modulation drive
to the other quadrature, namely $\phi$, and measure charge. The
modulation in the off-diagonal components would result in slightly
different interference phenomena, as predicted recently by
H\"anggi and coworkers \cite{wubs05}.

The coherent Landau-Zener process observed in this work also
serves as a suitable method, intermediate between the two existing
schemes, for manipulations of superconducting qubits. Thus far,
qubit operations have been carried out either using lengthy
rf-pulses which induce Rabi oscillations, or by sub-nanosecond
"hard" pulses~\cite{Nakamura99} shaped rectangularly in time. The
LZ manipulation would offer an ultra-short clock period $\ll$ ns
similarly to hard pulses, but would be more precise due to
basically single-frequency drive, and also technologically
simpler.

{\flushleft{\bf Acknowledgements}}\\
Fruitful discussions with R.~Fazio, M.~Feigelman, T.~Heikkil\"a,
F.~Hekking, and M.~Paalanen are gratefully acknowledged. This work
was financially supported by the Academy of Finland, by the
National Technology Agency, and by the Vaisala Foundation of the
Finnish Academy of Science and Letters.

\bigskip
{\flushleft{\bf Methods}}\\
Our investigations of the LZ tunneling are based on measuring the
quantum (or "Josephson") capacitance of a
CPB~\cite{Mika05,Duty05}. This capacitance is related to the
curvature of band $k$ \cite{widom84,Likharev85}, similar to the
effective mass of an electron in a crystal:
\begin{equation}
C_{\rm eff}^{(k)} = - \frac{\partial^2 E_k(\phi, n_g)}{\partial
V_g^2} = - \frac{C_g^2}{e^2} \frac{\partial^2 E_k (\phi,
n_g)}{\partial n_g^2}\,. \label{Ceff_defin}
\end{equation}
The difference in the Josephson capacitance for $k = 0,1$ allows
us to determine the state of the CPB.

We perform low-dissipation microwave reflection measurements
\cite{lset,leif05,MIKAthesis} on a series $LC$ resonator in which
the box effective capacitance, Eq.~(\ref{Ceff_defin}), is a part
of the total capacitance $C_S+C_{\mathrm{eff}}^{(k)}$. The
resonator is formed by a surface mount inductor of $L = 160$ nH.
With a stray capacitance of $C_S = 250$ fF due to the fairly big
lumped resonator, the resonant frequency is $f_0 = 800$ MHz and
the quality factor is $Q \simeq 16$ limited by the external $Z_0 =
50 \, \Omega$. When $C_{\mathrm{eff}}^{(k)}$ varies, the phase
$\arg (\Gamma)$ of the reflected signal $V_{\mathrm{out}}=\Gamma
V_{\mathrm{in}}$ changes, which is measured by the reflection
coefficient $\Gamma = (Z-Z_0)/(Z+Z_0)=\Gamma_0 e^{i \arg
(\Gamma)}$. Here, $Z$ is the resonator impedance as marked in
Fig.~\ref{scheme}. Since we work rather far from matching
conditions, the magnitude of the reflection coefficient $\Gamma_0$
remains always close to one. The variation in $\arg (\Gamma)$ due
to modulation in $C_{\mathrm{eff}}^{(k)}$ is up to 40$^{\circ}$ in
our measurements, corresponding to a shift of resonance frequency
$\Delta f_p \simeq 6$~MHz. In addition to band pass filtering, we
used two circulators at 20~mK to prevent the back-action noise of
our cryogenic low-noise amplifier from reaching the qubit. In all
the measurements, the probing signal $V_{\mathrm{in}}$ was
continuously applied.

\bibliography{lzs}

\linespread{1.5}

\begin{figure}
\center
 \includegraphics[width=17.0cm]{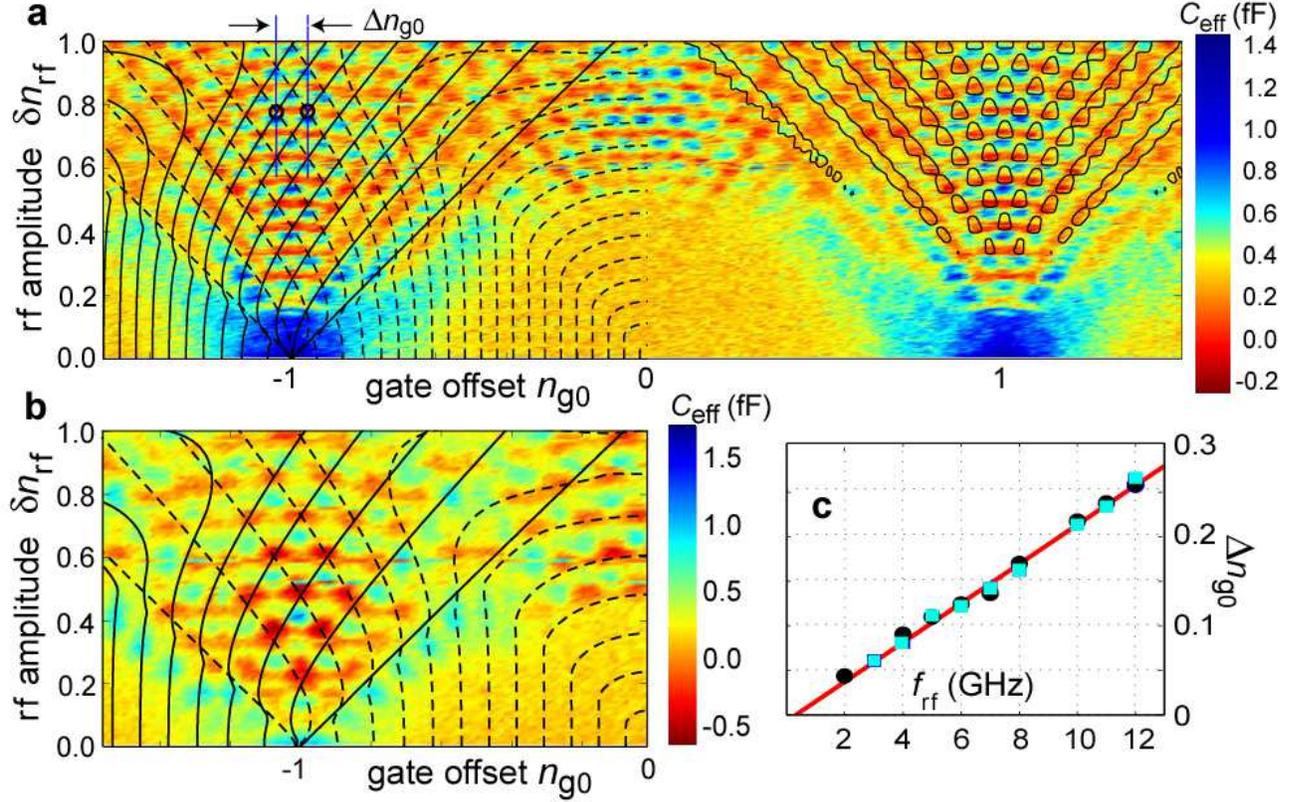}

\caption{{\large\textbf{a}}, interference pattern obtained from
the measured microwave phase shift at $f_{\mathrm{rf}} = 4$ GHz.
The phase bias $\phi = 0$, corresponding to the level repulsion of
$2 \Delta = 2 E_J = 12.5$ GHz. The color codes indicate the
equivalent capacitance obtained using standard circuit formulas
(see Methods). Around $n_{g0} = -1$, the imposed lines illustrate
the conditions of constructive Landau-Zener interference, which is
expected when the phases picked up to the left of $n_{g} = -1$,
$\varphi_L - \pi + 2 \phi_S$ (solid), and to the right, $\varphi_R
- \pi + 2 \phi_S$ (dashed), are integer multiples of $2 \pi$ (see
Eq.~(\ref{phasecondition}). Here, due to the almost adiabatic
limit, the Stokes phase $\phi_S = 0$. The population of the upper
state is expected to be the strongest (red) when both conditions
are satisfied. Around $n_g = 1$, is imposed the equicapacitance
contour for $C_\mathrm{eff} = 0$ fF obtained from the spin
dynamics simulation using Bloch equations (compare to
Fig.~\ref{trace}) which agrees quite well with both the resonance
grid as well as with the data. The interferogram
{\large\textbf{b}} is similar to \textbf{a} but with
$f_{\mathrm{rf}} = 7$ GHz, and $\phi_S = \pi/4$ due to operation
in almost the "sudden" limit of LZ tunneling. {\large\textbf{c}},
the averaged spacing of the central interference peaks in gate
offset (as depicted in \textbf{a}), with the phase bias values of
0 (squares) and $\pi$ (circles). The expected linear behavior
yields a fit $E_C = 1.1$ K.} \label{C3d4GHz}
\end{figure}

\begin{figure}
\center
 \includegraphics[width=15.0cm]{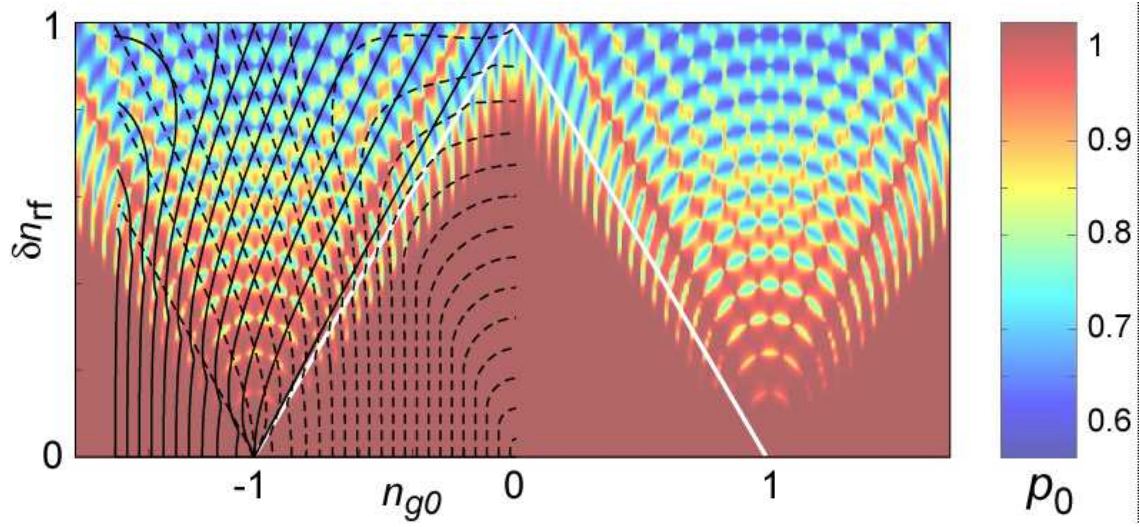}


\caption{Interference-induced variation in level populations
(color codes indicate relative population $p_0$ of the ground
state) obtained from the simulation of Bloch equations for the
qubit using a driving field of $B_z(t)=2E_C(1-n_{g0}+\delta
n_{\mathrm{rf}} \sin(\omega_{\mathrm{rf}}t))$, with
$\omega_{\mathrm{rf}}/(2\pi) = 4$ GHz. Ohmic bath was assumed for
the dissipation and the parameter $\alpha$ of Eq.~(\ref{t1}) was
set to 0.04. The inclined white lines illustrate the locations of
reaching the degeneracy point during the sweep for the first time
when ramping up $\delta n_{\mathrm{rf}}$ at fixed $ n_{g0}$
(starting point with $\delta n_{\mathrm{rf}}<1-n_{g0}$). The solid
and dashed black lines indicate the locations where conditions for
$2\pi$-multiple phase shifts on the both sides of the degeneracy
point are fulfilled, Eq.~(\ref{phasecondition}), with $\Phi_S = 0$
(similarly as in the experimental data in Fig.~\ref{C3d4GHz}).}
\label{alpha1}

\end{figure}

\begin{figure}

\center
 \includegraphics[width=13.0cm]{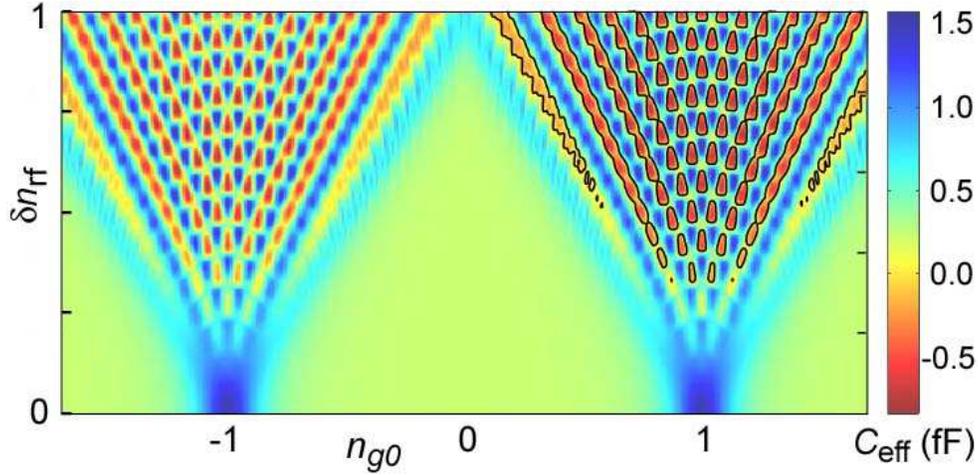}

\caption{Josephson capacitance $C_\mathrm{eff}$, calculated as
arising from curvatures of the two lowest bands and interband
relaxation, using Bloch equations and linear response theory. The
dissipation parameter $\alpha$ of Eq.~(\ref{t1}) was set to 0.04
and the amplitude of the ac-excitation at 803 MHz corresponds to
0.06e peak-to-peak. The capacitance variation is intermediate
between simple averaging and the fast relaxation approaches, and
it agrees quite well with the measured results in
Fig.~\ref{C3d4GHz}: the comparison with data is performed by the
displayed contour graph which describes the equicapacitance curve
for $C_\mathrm{eff} = 0$ fF obtained from the simulation.}
\label{trace}
\end{figure}

\end{document}